\documentclass[nofootinbib,prd,superscriptaddress,showpacs,showkeys,]{revtex4}

\usepackage{amsmath}
\usepackage{amssymb}
\usepackage{graphicx}
\usepackage{epstopdf}
\usepackage{inputenc}

\usepackage[usenames,dvipsnames]{color}
\usepackage{hyperref}

\newcommand*{\ket}[1]{\ensuremath{|#1\rangle}}
\newcommand*{\bra}[1]{\ensuremath{\langle#1|}}
\newcommand*{\bbN}{\ensuremath{\mathbb N}}

\begin{document}

\title{Information entropy and dark energy evolution}

\author{Salvatore Capozziello}
\email{capozzie@na.infn.it}
\affiliation{Dipartimento di Fisica, Universit\`a di Napoli ''Federico II'', Via Cinthia, I-80126, Napoli, Italy.}
\affiliation{Istituto Nazionale di Fisica Nucleare (INFN), Via Cinthia, I-80126, Napoli, Italy.}
\affiliation{Gran Sasso Science Institute, Viale F. Crispi, L'Aquila, Italy}

\author{Orlando Luongo}
\email{luongo@na.infn.it}
\affiliation{Dipartimento di Fisica, Universit\`a di Napoli ''Federico II'', Via Cinthia, I-80126, Napoli, Italy.}
\affiliation{Istituto Nazionale di Fisica Nucleare (INFN), Via Cinthia, I-80126, Napoli, Italy.}
\affiliation{Department of Mathematics and Applied Mathematics, University of Cape Town, Rondebosch, 7701, Cape Town, South Africa.}
\affiliation{School of Science and Technology, University of Camerino, I-62032, Camerino, Italy.}

\begin{abstract}
The information entropy is here investigated in the context of early and late cosmology under the hypothesis that distinct phases of universe evolution are entangled between them. The approach is based on the \emph{entangled state ansatz},  representing  a coarse-grained definition of primordial \emph{dark temperature} associated to an \emph{effective entangled energy density}. The dark temperature definition comes from assuming either Von Neumann or linear entropy as sources of cosmological thermodynamics. We interpret the involved information entropies by means of probabilities of forming structures during cosmic evolution. Following this recipe, we propose that  quantum entropy is simply associated to the thermodynamical entropy and we investigate the consequences of our approach using the adiabatic sound speed. As  byproducts, we analyze two phases of universe evolution: the late and early stages. To do so, we first recover that dark energy reduces to a pure cosmological constant, as zero-order entanglement contribution, and second that inflation is well-described  by means of an effective potential. In both cases, we infer numerical limits which are compatible with current observations.
\end{abstract}

\pacs{98.80.-k, 98.80.Qc, 03.67.-a}

\keywords{quantum cosmology; entanglement; decoherence; dark energy.}

\maketitle


\section{Introduction}\label{introduzione}

The issue of determining the universe evolution has brought physicists to formulate various approach spanning from modifications of the Einstein-Hilbert action, quintessence fields to barotropic fluids, and so forth \cite{rev1,rev1bis,rev2,rev2bis}. A recent wide interest has been devoted to stemming the universe dynamics from quantum effects, through a process named decoherence \cite{deco}. The problem of postulating a self consistent evolution of the net quantum fluid is associated to the lack of a complete quantum gravity approach \cite{a,patrizia}. All the possibilities, developed in the literature are typically overshadowed by several bugs. However, a broad interest has been pursued in order to reach suitable quantum signatures in observational cosmology. Indirect hints of plausible quantum shadows may be viewed in constraining the cosmological constant $\Lambda$ from cosmological observational bounds \cite{lambda}. Phrasing it differently, theoretical predictions of a gravitational vacuum state are essentially incompatible with cosmic observational limits. Moreover, both $\Lambda$ and matter orders of magnitude are surprisingly close to each other, albeit the cosmological constant does not evolve in time \cite{lambda2}. A way out for alleviating the above mentioned problems takes into account an additional classical fluid, dubbed dark energy associated to an evolving negative equation of state \cite{darkenergy}. This represents a basically classic approach since the fluid enters the Friedman equations postulating an additional contribution to the whole energy momentum tensor \cite{darkenergy2}. The possible signature of quantum effects at high redshift provides that dark energy may be viewed as a resulting effect of some quantum process \cite{ioesalv}. In such a way, the existence of a hidden quantum mechanism behind the nature of the cosmological constant may be of great interest to describe the dark energy nature \cite{quantumlambda}. A straightforward approach lies on applying the standard quantization to the dynamical equations, passing from the full infinite-dimensional configuration space of general relativity to the so called minisuperspace approach. In that case, the degrees of freedom are mostly reduced and the universe wave function 	 is determined up to boundary conditions, essentially related to the dynamical quantities which define the universe expansion history. Another relevant example is the study of the entanglement effects in primordial cosmology \cite{jae1,jae2}. They have been extensively investigated in the development of modern quantum information \cite{entareview}. The idea of entanglement strives with classical intuition, although recently increasing experimental tests have been performed. The corresponding excellent experimental outputs definitively forecast the prominent role played by entanglement, in various contexts, spanning from quantum physics to cosmology \cite{entareview2}.

In this work, we basically ask for  quantum correlations that can emerge   in  the framework of  cosmology. We assume that quantum information entropy may enter the  energy-momentum tensor as source for the universe dynamics \cite{cosmoenta}. Hence, it turns out that primordial entangled states may contribute with a non-zero density matrix whose trace enters the net energy density, driving the cosmic speed up at different stages of the universe evolution. While at late times the cosmological constant is essentially recovered as limiting case, at higher redshift it becomes a consequence of the adiabatic exchange between cosmological components. The dark energy contribution can be viewed, in general, as an evolving term obtained  as an effect of quantum information \cite{cosmoenta2}. To demonstrate this fact, we here first build up the Von Neumann entropy as a source for the cosmological entropy \cite{cosmoenta3}. Our strategy is to take the Von Neumann entropy between two entangled states, which correspond to two different epochs of the universe evolution. Afterwards, we infer the corresponding dark energy temperature and  interpret it as the quantum information temperature, in the special case of homogeneous and isotropic universe. In particular, specific classes of dark energy models  are  considered, scaling the density term as pure matter, dark energy and the sum between them. In so doing, we  show that the entanglement mechanism is applicable  to the  degrees of freedom  of macroscopic systems, for the whole  universe evolution, under the standard  of decoherence \cite{sett}. This effect, in the cosmological puzzle, is due to the interaction between {\it qubits} with their environment \cite{sett2}, which is the universe itself.

In the paper, we study the density, pressure and equation of state associated to the dark temperature and  compare  results with the  linear entropy, as limiting case of the Von Neumann entropy. Moreover, we consider the adiabatic hypothesis to define either the adiabatic index or the sound speed. From those theoretical constraints, we set stringent limits on the cosmological  parameters. We find that the model works fairly well in reproducing both early and late time cosmological  evolution. Indeed, the cosmological constant is recovered as a limiting case and is interpreted as zero-value of the evolving information density, entering the energy-momentum tensor. Analogously, the  approach can be adopted for early stages and one demonstrates that inflation is also recovered if an effective field is defined from entanglement. We show that the numerical results inferred at late times are perfectly compatible in determining the spectral index of inflation and turn out to reproduce the inflationary accelerated scenario.

The  layout of the paper  is the  following. In Sec. II, we discuss  quantum states that can be built up in view of  the Von Neumann entropy. In particular, we take into account the role played by the entanglement, using  two quantum states in  homogeneous and isotropic cosmology. Afterwards, we combine these requirements with the standard Friedman equations. In Sec. III the role of  dark temperature and its evolution is considered giving a statistical explanation to the Von Neumann entropy by means of associated probabilities of structure formations. In Sec. IV, the case of adiabatic expansion is reviewed with particular emphasis for the consequences in late and early time cosmology. Conclusions are drawn in Sec. V.


\section{Building up entangled states for cosmological eras}

Let us  introduce a strategy for characterizing cosmological quantum states and for describing \emph{a posteriori}  entanglement effects between them.  In view of this, a heuristic procedure consists in  choosing the minimal number of quantities capable of describing the universe at different stages of its evolution according to  the cosmological principle. Thus, a possible
\textit{vector field} obtained by observables is:
\begin{widetext}
\begin{eqnarray}\label{setto}
\boxed{\left\{\Theta\right\}\,=\,\left\{H(z), q(z), \Omega_{m}(z), \Omega_{k}(z), T(z), \ldots\right\}\,,}
\end{eqnarray}
\end{widetext}
where  in the  brackets there are  \emph{cosmographic parameters}, i.e. quantities that can be, in principle, measured without imposing   \emph{a priori}  a cosmological model\cite{cosmografia}. They are the Hubble parameter $H$, the {\it deceleration} parameter, the matter density $\Omega_{m}$  and  the spatial curvature density  $\Omega_{k}$ respectively\footnote{Clearly, the "dimensions" of the vector (\ref{setto}) can be improved according to the degree of "refinement" of our considerations: less parameters mean coarse-grained models.} \cite{cosmografia2}. We include also $T$,  the cosmic  temperature  following from the definition of cosmic  entropy \cite{entropiaetemperatura}.
 Let us assume now now some prescriptions  that we are going to adopt throughout the paper.

\begin{enumerate}
  \item Provided two distinct epochs of the universe evolution, a plausible state which involves a two dimensional vector defined over a given complex Hilbert space $\mathbb{C}$ is:
\begin{equation}\label{phi}
 |\phi\rangle\equiv \left(
\Omega_{m}+i\Omega_{k},\\
\Omega_{k}+i\Omega_{m}\right)^T,
\end{equation}
where the observable quantities  are the spatial curvature ($\Omega_{k}$) and matter ($\Omega_m$) at those epochs.

  \item Following the basic requirements of quantum mechanics,  to get a linear independent and non-normalized vector in the  Hilbert space, we can adopt the following construction:
\begin{eqnarray}
|\tilde e_A\rangle&=&\left(
\Omega_{m}+i\Omega_{k},
i\Omega_{m}+\Omega_{k}\right)^T, \\
|\tilde e_B\rangle&=&\left(
\Omega_{m}-i\Omega_{k},
-i\Omega_{m}+\Omega_{k}\right)^T.
\end{eqnarray}

  \item Related to the above prescription, one can assume the validity of the Gram-Schmidt procedure which turns out to give orthonormal bases of the form:
\begin{eqnarray}
|e_A\rangle&=&
N_A |\tilde e_A\rangle,\label{ortho1}\\
|e_B\rangle&=&
N_B (\Omega_m^2+\Omega_k^2) |\tilde e_B\rangle
+N_B 2i\Omega_m\Omega_k |\tilde e_A\rangle,
\label{ortho2}
\end{eqnarray}
with normalization factors: $N_A=\frac{1}{\sqrt{2(\Omega_m^2+\Omega_k^2)}}$ and $N_B=\frac{1}{\sqrt{2(\Omega_m^2+\Omega_k^2)(\Omega_k^2-\Omega_m^2)^2}}$.

\item For any epochs, e.g. inflation, reheating, recombination, dark energy domination and so forth,  there would exist a different Hilbert space, which is built up guaranteeing that the cosmological principle holds. To account for the cosmic dynamics we assume that, as the universe undergoes a phase transition, the \emph{interaction} of any phase is modeled by the entanglement between quantum states. Hence,   entanglement replaces the effects due to phase transitions in the standard cosmology picture.

\item  As a result, we infer a corresponding \emph{multipartite} system. To allow the entanglement process, in particular, we need to consider at least two epochs, leading to an associated Hilbert space of the form: $\mathbb{C}^2\otimes \mathbb{C}^2$, whose orthonormal basis is given by:
\begin{eqnarray}
 |e_A\rangle_{1} |e_A\rangle_{2},\;
  |e_A\rangle_{1} |e_B\rangle_{2},\;
   |e_B\rangle_{1} |e_A\rangle_{2},\;
    |e_B\rangle_{1} |e_B\rangle_{2},
 \label{basis}
\end{eqnarray}
with $1;2$ indicating the epochs involved.
\end{enumerate}
The cosmic state is univocally assigned to a certain time of cosmological evolution as one defines the functional dependence of the variables associated to that state. If the cosmic \emph{evolution} is due to the correlation among states, with constraints compatible with present observations, we soon get that any coarse-grained quantum state is modeled by means of  the vector  \eqref{setto}. The principal consequence of assuming a vector of such a form is that the corresponding thermodynamical properties emerge from the \emph{information entropies} \cite{nuew}, defined as byproducts of the  entangled wave functions. This procedure does not need \emph{a priori} a process of quantization of gravity
\cite{nicroro2,nicroro3}, requiring, instead, that the universe is characterized by a quantum description at the level of primordial cosmology \cite{queb}.

The entangled \emph{correlation} is hereafter named \emph{entangled states ansatz}, where the two epochs are built up in terms of the following wave function:
\begin{equation}
|\Psi\rangle = \alpha|e_A\rangle_1|e_B\rangle_2 + \beta|e_B\rangle_1|e_A\rangle_2\,,
\label{stato}
\end{equation}
where, $\alpha,\beta\in\mathbb{C}$ lead to $|\alpha|^2+|\beta|^2=1$, corresponding to the standard normalization procedure.

The aim of this paper is to start from the above assumptions and to describe  quantum thermodynamics in  cosmology. This means to find  effective relations between the Von Neumann entropy and cosmological densities in view   of  dark energy evolution stemming out from early epochs.


\subsection{The early phases of cosmic evolution}

The strategy to match entanglement with early universe thermodynamics consists in comparing thermodynamical potentials with quantum entropies. To do so, let us take into account the above described $N$-dimensional Hilbert space and let us associate to a certain event a given probability $\mathcal P_k = \frac{1}{N}$, $\forall k$. We require that a particular observable is associated to $\mathcal P_k$ and even that $\mathcal P_k\neq 0$ for entangled quantum states only.

Thus, considering non-local entangled states with $\mathcal P_k\neq 0$, we associate any probability to the cosmological quantities of  interest, for example distribution of  galaxies, cosmic structures, and so on. For those reasons, \emph{selection criteria}, i.e. techniques to quantify \emph{how much} a given cosmological quantum state is actually non-pure, become essential to characterize the \emph{cosmological density matrix}.

Hence, specializing our attention to two distinct epochs in which the universe is described by a set of observable quantities, we associate to them two quantum states. Those states depend upon a finite number of cosmological quantities which  describe the universe dynamics even at a classical regime. Phrasing it differently, one looks for the possibility to define the set of parameters,  which enable either a classical or quantum descriptions of the whole universe, into a single paradigm.

Under the hypothesis of a homogeneous and isotropic space-time, the simplest assumption is to take into account the matter density, constituted by baryons and cold dark matter, and the spatial curvature only. So that, considering the Friedman-Robertson-Walker (FRW) metric:
\begin{equation}\label{hgjhg}
ds^2=dt^2-a(t)^2\left(dr^2+r^2\sin^{2}\theta d\phi^2+r^2d\theta^2\right)\,,
\end{equation}
we assume that all fluids entering the standard energy momentum tensor are perfect. Thus, the  equation of state for any species is defined by $\omega_i\equiv\frac{P_i}{\rho_i}$, whereas the pressure for the dark component must exhibit repulsive effects in order to counterbalance the attraction of gravity \cite{iorepulsivo,ionorepulsivo}. Considering the conservation law given by the contracted Bianchi identities $\nabla^\alpha T_{\alpha\beta}=0$, one gets the relation
\begin{equation}\label{utyyyy}
\frac{d\rho}{dz}=3\left(\frac{1+\omega}{1+z}\right)\rho\,,
\end{equation}
where $\omega$ is the adiabatic index of tjhe equation of state and $z$ is the redshift.
The Friedman equations are

\begin{eqnarray}
\label{ave2}
H^2 &=& {8\pi G\over3}\rho\,,\\
\dot H + H^2&=&-{4\pi
G\over3}\left(\rho+3P\right)\,,\nonumber
\end{eqnarray}
characterizing the universe dynamics at all stages of  evolution. Here, we assume the Hubble parameter definition $H=\dot a/a$ and the spatial curvature to be negligibly small\footnote{This assumption does not contradict the above construction of quantum states, but  simply reduces the dimensions of parameter phase space   simplifying  the form of wave function.}. Associated to any species, one  has a corresponding temperature as a  function of $\omega_i$ or, more practically,  as function of $z$. Even though the \emph{dark temperature}, i.e. the temperature of dark energy and dark matter, may be, in principle, the dominant term for some epochs of the universe, no significative  modifications are today expected as provided by observations.

To match  the  quantum state description, given by Eq. \eqref{stato}, with the classical description related to   Eqs. \eqref{ave2}, we can easily assume that the density matrix is an \emph{energy source} for $\rho$. This can be obtained by assuming that the eigenvalues of the density operators are the \emph{present observed densities}.

We will  use of the above recipe  to find out an expression for $P, \omega$ and $\rho$ in the case in which the dark energy term is derived from quantum effects. The idea is supposing that the Von Neumann entropy is a suitable measure for entanglement  and contributes to the whole entropy of the universe, assuming that the eigenvalues of the density operator enter Eqs. \eqref{ave2}. Assuming the Von Neumann entropy as density source for the Friedman equations is not a mere imposition. It is possible to  demonstrate, indeed,  that under the hypothesis of binomial probability distribution, one recovers the Von Neumann entropy as an entropy source. This fact enable us to describe the universe through a multipartite state.


\section{The quantum \emph{dark temperature}}

We focus now on the consequences on thermodynamics related to  Eq.\eqref{stato}. In particular, let us  redefine the concept of temperature and  discuss the role of the Von Neumann entropy in the cosmic evolution. In particular, we have to note that microphysics of dark energy is still unknown due to the fact that a  self-consistent theory of quantum gravity is currently unavailable. Plausibly, a full quantum theory of gravitation would provide a final description of the \emph{effective} dark energy evolution, describing the reasons for a too large quantum cosmological constant contribution if compared with cosmic observations. A description  based on  \emph{decoherence} can help in this perspective. Decoherence allows the emergence of classical features  when some physical degrees of freedom interact with the quantum system by the phenomenon of entanglement, i.e. some  degrees of freedom entangle with the environment. Our prescription is that different epochs of the universe evolution provide some physical degrees of freedom which are intimately interacting with the universe environment. In other words, one expects that  macroscopic physical properties  can emerge from this process. In particular, thermodynamical properties.
Combining the first and second laws with $V=a^3$ the volume, we  obtain the relations for the entropy
\begin{subequations}
\begin{align}
Td\mathcal S&=d(\rho V)+P dV = d[(\rho+P)V]-V dP\,,\\
{\partial^2 \mathcal S \over \partial T \partial V}&={\partial^2 \mathcal S \over \partial V \partial T}\,.
\end{align}
\end{subequations}
The second relation comes from the  integrability. Immediately we have
\begin{equation}
dP=\left({\rho+P\over T}\right)dT\,,
\end{equation}
whose form can be integrated to give:
\begin{equation}
d\mathcal S= {1\over T} d[(\rho + P)V]-(\rho + P)V {dT\over
T^2}=d\bigg[{(\rho+P)V\over T} + C \bigg]\,,
\end{equation}
where  $C$ is an integration constant. The entropy density naturally follows as
\begin{equation}\label{entrotermo}
\mathcal S \equiv{\rho + P\over T}\,.
\end{equation}
The above calculations are standard. We need now a mechanism such that classical aspects emerge as consequence of quantum evolution. In other words, we need to relate Eq. \eqref{entrotermo} to the above quantum description. One can assume the  process of decoherence which supposes that the universe is localized in a given  state  defined in one of the minima of an effective potential. In one of these minima, for example, the universe can start the today accelerating phase. In particular, during the evolution in such a state the  equation of state persists in its negative form, whereas afterwards it changes its sign. This happens because there exists a given probability that the universe evolution falls into a superposition of localized states, which correspond to a ground state. The assumption of different vacua can be replaced considering entanglement as responsible for the same process.

\subsection{The  Von Neumann entropy at early time cosmology}

In order to motivate how Eq. \eqref{entrotermo} is fueled by quantum entropy, we relate $\mathcal S$ to its quantum counterpart. To get the form of quantum information entropy, let us assume that a cosmological structure, at a given epoch, forms with a  probability $\mathcal P_1$ given by:
\begin{subequations}\label{prob1}
\begin{align}
\mathcal P_2&=1-\mathcal P_1\,,\\
\mathcal P_1&=\rho_n\,,
\end{align}
\end{subequations}
where $\rho_n$ is the \emph{normalized} density of microstates associated to the system under examination, i.e. the set of eigenvalues of the density operator. Under this scheme, the probability of non-forming a given galaxy can be reported as $P_2$.
If one assumes that the net information contribution  $\mathcal I$ for structure formation must be positive definite, a possible relation with the entropy is $I\propto \exp S$, where $S$ is the information entropy that (in principle) is different from $\mathcal S$. In so doing, one gets:
\begin{equation}\label{entropy32}
S=\ln\rho_n\ln(1-\rho_n)\,,
\end{equation}
which comes from  the binomial distribution:
\begin{equation}\label{bi}
P(X)={N\choose X}\rho_n^N(1-\rho_n)^{N-X}\,,
\end{equation}
with the ansatz that $S\propto \mathcal P(X)$ and $N=2, X=1$.
If the density of microstates is small enough, Eq. \eqref{entropy32} becomes:
\begin{equation}\label{entropy33}
S\propto-\rho_n\ln\rho_n\,,
\end{equation}
which is much more accurate as soon as $\rho_n\rightarrow0$.
Since the definition \eqref{entropy33} is general, one can reverse the approach and may imagine that the Von Neumann contribution well adapts to Eq. \eqref{entropy33}. Thus, if we assume that the standard entropy density, evaluated by Eq. \eqref{entrotermo}, is entirely due to the information entropy, one can conclude that
\begin{equation}\label{uguaglianza}
\mathcal S\propto \exp I =S\,.
\end{equation}
In this picture, the cosmic entropy comes out from   quantum contributions without considering  any decoherence process. In what follows, we therefore assume the density entering Eqs. \eqref{ave2} is given by $\rho\equiv\sum_n \rho_n$.

\subsection{From information to quantum entropy}

An immediate extension of what we described in terms of probability is based on introducing the \emph{reduced density matrix} $\hat{\rho}$. A mixed configuration  is
\begin{equation}\label{mixed}
\text{Tr}\hat{\rho}^2 \neq \text{Tr} \hat{\rho}\,,
\end{equation}
so that we can define the $m$-particle reduced density matrices by:
\begin{equation}\label{e:yy}
  \rho_m \equiv \sigma\ket{i}\text{Tr}\{a_\star\,\,
  \rho\,\,a^\dag_\star \}\bra{j}.
\end{equation}
where $a_\star\equiv \left\{a_{i_1}, a_{i_2},...,a_{i_m}\right\}$, $a_\star^\dag\equiv \left\{a_{i_1}^\dag, a_{i_2}^\dag, ..., a_{i_m}^\dag\right\}$, $\ket{i}\equiv\ket{i_1, i_2, ..., i_m}$ and $\bra{j}\equiv \bra{j_1, j_{2}, ..., j_m}$, with $\sigma\equiv \frac{1}{(m!)^2}$.
An important property is that,  under the above  reduced density matrices unitarily transform being $\rho' = U(\Lambda)\rho U^\dag(\Lambda)$.  This means that they are  invariant under Lorentz transformations.
In other words,  taking into account a  Lorentz transformation $\Lambda$, all quantum operators and in particular creation and annihilation operators, transform according to  standard Lorentz transformation. Moreover:
\begin{eqnarray}
\mathcal U(\Lambda)\ket{0}=\ket{0}\,,
\end{eqnarray}
where the last condition leads to the Lorentz invariant vacuum and allows to determine all states acting as creation and annihilation operators on it. This holds if and only if the vacuum unitarily transforms via the relations: $ \ket{\Psi}' = \mathcal U(\Lambda)\ket{\Psi}, \bra{\Psi}'=\bra{\Psi}U^\dag(\Lambda)$.

\subsection{The Von Neumann entropy and its consequence in cosmology}

From the above considerations, the von Neumann entropy may be considered as a natural consequence of the information \emph{broadcasted} by the density matrix. We follow this recipe to adapt the Von Neumann entropy in the framework of primordial cosmology. To do so, we basically adopt the idea: \emph{if a quantum (cosmological) state exists under the hypothesis to be a non-pure state, then it shows an entropy of the form}:
\begin{equation}
S \equiv -\text{Tr}{(\hat{\rho}\ln\hat{\rho})} = -\sum^N_{n=1} \rho_n\ln\rho_n.
\end{equation}
Here, $\rho_n$ are the eigenvalues of $\rho$, with the additional normalization condition $
\text{Tr}\hat{\rho} = 1$.
The density matrix is commonly associated to physical states, so that it should be Hermitian, with the property:
\begin{equation}\label{hermitiano}
\rho \circeq V[diag({\bf \rho}_n)]V^{\dagger}\,,
\end{equation}
which implies that the eigenvalues ${\bf \rho}_n\equiv\rho_1,..., \rho_{N}$ are invariant under unitary transformations.
The same property is accounted for Lorentz transformations $\mathcal U(\Lambda)$. In this case, we require that
\begin{eqnarray}\label{lorentz}
\hat{\rho}^\prime &= \mathcal U(\Lambda)\rho
\mathcal U^\dag(\Lambda) &= \mathcal U(\Lambda)V[diag({\bf \rho}_n)][{\mathcal U(\Lambda)V}]^{\dagger}\,,
\end{eqnarray}
which preserves the eigenvalue values, providing the above  property that the von Neumann entropy turns out to be \emph{Lorentz invariant} and therefore \emph{it can be used in  cosmology}. As well as $\hat\rho$, even the Von Neumann entropy of every \emph{reduced density matrices}, i.e. $S_r\equiv -\text{Tr}{(\rho_R\ln\rho_R)}$, with $R\in[1;{\bbN}]$,  satisfies the Lorentz invariance. The universe can be therefore  parameterized by means of $\hat\rho$. The strategy lies on assuming a set of Lorentz invariant (separate) measurements for arbitrarily chosen epochs of its evolution.

Hence, parameterizing by the Von Neumann entropy,   assuming  $\mathcal S=S$,  and inverting Eq. \eqref{entrotermo}, we have:
\begin{equation}\label{temp}
T=-\frac{P+\rho}{\rho\ln\rho}\,,
\end{equation}
with the request that the sum of eigenvalues gives rise to $\rho$. Here $T$ is the dark temperature. Eq. \eqref{temp} can be recast in terms of  redshift, by means of Eq. \eqref{utyyyy}:
\begin{equation}\label{tempz}
T=-\frac{(1+z)}{3\rho\ln\rho}\frac{d\rho}{dz}\,,
\end{equation}
or alternatively, in a more compact form, adopting the identity $a\equiv(1+z)^{-1}$ by:
\begin{equation}\label{tempcompact}
\boxed{T=\frac{d\ln\Big[\ln\rho\Big]^{\frac{1}{3}}}{d\ln a}\,.}
\end{equation}
Measurements of cosmic microwave background radiation allow to constrain the temperature of the universe. The temperature involved in the standard measure is considered as due to all constituents of the universe. Thus, our outcome enters the net temperature as well, contributing with a term of dark temperature, i.e. associated to quantum imprints at early stages.

The functional behavior of $T$ in terms of $z$ is not known \emph{a priori}. We may suppose to take a generic polynomial expansion of $T$. In particular, the simplest assumption can be $T\propto a^n$, giving:
\begin{subequations}
\label{soluz}
\begin{align}
\rho&=\rho_C \exp(f_3(a))\,,\\
\rho_C&=(1-\rho_0)f_3(1)\,,
\end{align}
\end{subequations}
where we made use of the function:
\begin{equation}\label{f}
f_\sigma(a)\equiv\exp\left({\sigma\,a^n\over n}\right)\,.
\end{equation}

\begin{figure}[ht]
\includegraphics[scale=0.6]{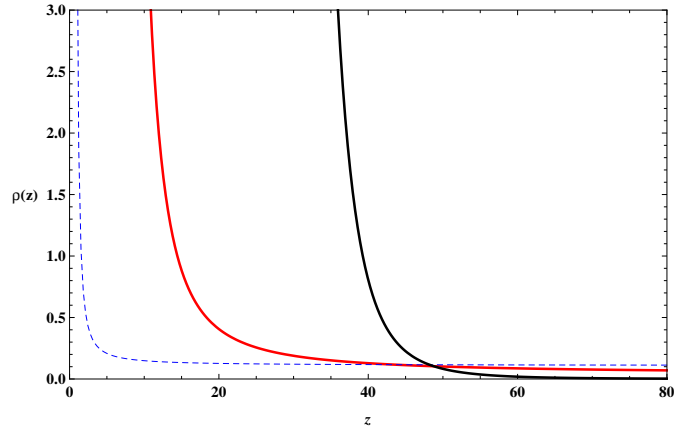}
\caption{{\it{The evolution of the function $\rho(z)$; the qualitative values here reported are $\rho_C=10^2; 25; 10^6$ for $n={1\over 2};\,1;\,{1\over 3}$ respectively red, dashed and black lines. }}}
\label{figura2}
\end{figure}

It could be convenient to focus on a first-order expansion leading to the approximate solutions:
\begin{subequations}\label{appr}
\begin{align}
\rho^{(1)}&\approx (1-\rho_0)(1-3f_2(1)z)\,,\\
T^{(0)}&\approx\Big|\frac{\log f_3(1)}{\log(1-\rho_0)}\Big|\,,
\end{align}
\end{subequations}
where we use  $a\equiv(1+z)^{-1}$ and  consider that $\rho$, at $a=1$, is forced to be $\rho(a=1)=1-\rho_0$. This is an arbitrary choice which enables one to fix $\rho(a=1)$ with different values, corresponding to the magnitudes imposed from observations. Examples are: $\rho(a=1)=1$ (dark energy and  matter contribution); $\rho_0=0$, $\rho(a=1)=0.7$ (dark energy contribution); $\rho=0.3$, $\rho(a=1)=0.3$ (matter contribution);  $\rho_0=0.7$ and so forth. This specializes $\rho_0$ to be the counterpart of our quantum dark fluid and gives to the approach the role of additional fluid entering the  energy-momentum tensor.

Under the above simple hypothesis, it is easy to show that  approximate ranges of values for $n$ are given by:
\begin{subequations}\label{enne}
\begin{align}
n&\in[-0.255\,;\,0[\,,\\
n&\in[-0.295\,;\,0[\,,
\end{align}
\end{subequations}
which guarantee that the paradigm gives rise to  pure dark energy (first case),  pressureless matter (second case), as a byproduct of the entangled state ansatz.

Another particularly interesting case is to consider as entropy source the \emph{linear entropy}. It represents the first prototype of entanglement entropy and may be viewed as a first approximation towards the definition of the Von Neumann entropy. In such a case, we have:
\begin{equation}\label{linearentropy}
S_L=\frac{N}{N-1}\left(1-\mu\left[\rho\right]\right)\,.
\end{equation}
Here, we can interpret $\mu[\rho]$ as a general function of $\rho$, that can be approximated to the first-order Taylor expansion, having $\mu\sim\rho$. The corresponding entropy function estimates quantum correlations, given a set of subsystems in the same way of Von Neumann entropy.
In this case, for the dark temperature, we  have:
\begin{equation}\label{temp}
T=\frac{a}{3\mathcal N}\frac{d\ln(1-\rho)}{da}\,,
\end{equation}
or alternatively, with respect to the redshift $z$:
\begin{equation}\label{tempz}
T=-\frac{(1+z)}{3\mathcal N}\frac{d\ln(1-\rho)}{dz}\,,
\end{equation}
where we consider, in both the approaches, the re-scaling constant $\mathcal N\equiv \frac{N}{N-1}$, which depends upon the degrees of freedom $N$ of the system. Following the same procedure that we used for the Von Neumann entropy, and assuming for example $N=4$, we  get:

\begin{subequations}\label{soluzu2}
\begin{align}
\rho&= 1 + \rho_C\exp{\Big[\frac{4\,(1+z)^{-n}}{n}\Big]}\,,\\
\rho_C&=-\rho_0\exp{\left(-\frac{4}{n}\right)}\,,
\end{align}
\end{subequations}
with the reported limits over $n$, Eqs. \eqref{enne}, which still continue holding. Even in this case, we find that $\rho$ can be only matter, dark energy or both, by tuning the solution through the free constant $\rho_0$.


\section{The  adiabatic expansion }

Standard thermodynamics assumes that, provided a particle ensemble, the dynamical equilibrium corresponds to a precise maximum of entropy. At early stages of  evolution, the thermal equilibrium is reached. Indeed, one usually assumes that above $100$ GeV, the  distributions are  the standard Fermi-Dirac and Bose-Einstein ones
\begin{subequations}\label{distribuzione}
\begin{align}
f_{BE}(p)=\frac{1}{e^{(\epsilon-\mu)/T}-1}\,,\\
f_{FD}(p)=\frac{1}{e^{(\epsilon-\mu)/T}+1}\,.
\end{align}
\end{subequations}
The entropy definition can be generalized as
\begin{equation}\label{entrgen}
dS=\frac{dU+pfV-\mu dN}{T}\,,
\end{equation}
by adding also the chemical potential $\mu$. From the second law of thermodynamics, the chemical equilibrium is reached as soon as the whole initial chemical potentials are summed up becoming equivalent to the net final ones. In this case, through Eq. \eqref{distribuzione}, given the mass and the momentum $m$ and $p$ respectively, for non-relativistic particles, we consider the energy $\epsilon(p)\sim m+\frac{p^2}{2m}$ and we easily get, for $n$ identical terms:
\begin{equation}\label{rhoux}
\rho=mn+\frac{3}{2}nT\,,
\end{equation}
supposing that $\epsilon(p)\propto \rho$. In particular, during early phases, neglecting the particle chemical potentials, one arrives to the definitions of pressure, density and number of particles:

\begin{subequations}\label{definizioni}
\begin{align}
P=\frac{g}{(2\pi)^3}\int{d^3pf(p)\frac{p^2}{3\epsilon}}\,,\\
\rho=\frac{g}{(2\pi)^3}\int{d^3pf(p)\epsilon(p)}\,,\\
n=\frac{g}{(2\pi)^3}\int{d^3pf(p)}\,,
\end{align}
\end{subequations}
which are valid if  the equilibrium hypothesis holds.

The  expressions in Eqs. \eqref{definizioni} are evaluated under the hypothesis of adiabatic regime where the  entropy is forced to be a constant, hereafter $\bar S$. By means of the fact that $dS=0\Leftrightarrow S=\bar S$, we  can invert the expression for the entropy obtaining
\begin{equation}\label{prevst}
P=\bar S T-\rho\,,
\end{equation}
and, being  $\rho=-\frac{S}{\ln\rho}$, we  have:
\begin{equation}\label{obt}
P=\bar S\left[T+(\ln\rho)^{-1}\right]\,.
\end{equation}
The above formula provides two relevant limits:
\begin{description}
  \item[At early phases] $P\,\sim T$\,,\\
  \item[At late phases] $P\,\sim \frac{1}{\ln\rho}$\,,
\end{description}
which correspond respectively to a pressure which, at late times depends essentially upon the density in a non-linear way. as a consequence,  evaluating the equation of state $\omega\equiv P/\rho$, we have:
\begin{equation}\label{bd}
\omega\equiv-\frac{\bar S}{\rho\ln\rho}\,.
\end{equation}
In particular, since $\bar S\equiv -\rho\ln\rho$,  it follows that $\omega=-1$ and we recover a   cosmological constant contribution.

In other words, our paradigm suggests that the cosmological constant exists even at the level of early phases as consequence of the entangled state ansatz, albeit it evolves in time throughout the universe expansion history, leading to a dark energy evolving time at small redshift regime.  In other words, \textit{the cosmological constant is the consequence of the adiabatic evolution}.

Assuming a  perfect fluid,  the corresponding speed of sound is \cite{octo}:
\begin{equation}\label{speedofsound}
c_s^2 = \left(\frac{\partial P}{\partial \rho}\right)\,.
\end{equation}
It must be compatible with small density perturbations, which barely propagate in the universe \cite{sound}. Hence, the speed of sound may be supposed to be negligibly small at all stages of the universe evolution, having $c_s= \sqrt{\gamma (C_p-C_V)T}\sim0$,  so that we infer:
\begin{equation}\label{adyab}
c_s^2=\bar S\Big[\frac{\partial T}{\partial\rho}+\frac{\rho}{\bar{S}^2}\Big]\,.
\end{equation}
At late times,  the entropy is large enough to enable the speed of sound to vanish.  In other words,the approach points out that  entropy is proportional to the square of the
velocity and temperature gradients. One can find analogous results even in other effective thermodynamical models, e.g.  in the case of ideal-gas-like universe with zero speed of sound or in   geometrothermodynamics \cite{dgeometro}.


\subsection{The cosmic evolution at  late-time dark energy}

Without  fixing \emph{a priori} the value of $\rho_0$, the pressure associated to  dark component becomes:

\begin{equation}\label{prex}
P=-\frac{(1-\rho_0)\exp(f_3(a)-f_3(1))(f_3(a)+a^{-n})}{a^{-n}}\,.
\end{equation}
This expression for the pressure ensures that  $z$ lies inside the observational intervals and generalizes the case where the entropy has been taken constant. The corresponding value of $P$ tends to $P\sim -0.7$, in the case of $\rho_0=0.3$. This means that dark energy can  be obtained by our quantum fluid as a dark fluid. The functional form of the pressure slightly departs from a constant value, leading to a possible constant dark energy term, different from a pure cosmological constant.

The corresponding functional form of the pressure $P$ is slightly evolving and turns out to be:
\begin{equation}\label{prex}
P\approx -\rho_\Lambda\,,
\end{equation}
in a wide range of redshifts, compatible with those where dark energy dominates. Here, we can assume $\rho_\Lambda\equiv \rho_0-1$. This result  gives an important outcome: from our model it is possible to recover the cosmological constant behavior in an effective semiclassical approximation. The $\Lambda$CDM model can be therefore mimicked by quantum effects due to entanglement with a dark energy fluid which naturally has a negative sign for the  dark energy pressure. In other words,  the $\Lambda$ term can be recovered through the  information entropy.


\subsection{The cosmic evolution at early inflationary phase}

The present approach can be adopted also  at inflationary epoch. The prescription assumes, as  source of inflation, the energy density \eqref{soluz} derived from entanglement. To show this, one can evaluate the corresponding energy density  by defining an effective potential which depends on a scalar field $\phi=\phi(a)$.  We interpret the inflaton field as an effective byproduct of the entangled state ansatz. So, assuming a generic primordial scalar field as source for the inflationary phase, we have

\begin{subequations}\label{assunzioninflazionarie}
\begin{align}
a&(t)\propto t^s\,,\label{uno1}\\
\rho&=\frac{1}{2}\dot \phi^2+V(\phi)\,,\label{uno2}\\
P&=\frac{1}{2}\dot \phi^2-V(\phi)\,,\label{uno3}\\
r&=16\epsilon\,,\label{uno5}\\
n_s&=1+2(\eta-3\epsilon)\,,\label{uno4}
\end{align}
\end{subequations}
with $s>1$, as in the case of standard inflation and $n_s$ the usual spectral index, with $\epsilon\propto \left(\frac{V^\prime}{V}\right)^2$ and $\eta\propto \frac{V^{''}}{V}$ \cite{inflazione1}. Inflation occurs if  Eqs. \eqref{assunzioninflazionarie} are capable of rapidly accelerating the universe. We thus notice that those quantities enter the amplitude of the primordial power spectrum \cite{inflazione2} by:
\begin{eqnarray}\label{uno6}
\Delta^2_{\cal R}=\frac{1}{8}\left(\frac{\kappa }{\pi \sqrt{\epsilon}}\right)^2H^2\,.
\end{eqnarray}
Thus, by comparing Eqs. \eqref{assunzioninflazionarie} with the prescriptions of our model, we get as effective potential:
\begin{equation}\label{qualcosa}
V(a)=\frac{(1-\rho_0)}{2}\exp(f_3(a)-f_3(1))\left(2+a^nf_3(a)\right)\,,
\end{equation}
which has been evaluated as a  function of the scale factor $a(t)$ and corresponds to $\rho-P$. The associated numerical limits on the spectral index $n_s$ are:
\begin{equation}
n_s\in[0.9968,1.0074]\,.\label{uno7}
\end{equation}
The above values have  been defined through the intervals given in Eqs. \eqref{enne}, by fixing $s\approx 1.3$. In particular, the maximum $n_s$ has been found in the case: $n\approx -10^4$, whereas the minimum $n_s$ as: $n_s=-0.295$. Both the values are derived at  $t\approx 10^2\,t_{Pl}$, with $t_{Pl}$ the Planck time and conventionally $\rho_0=0.7$.

Although only indicative, those values may be directly compared with the observational values of $n_s$, estimated by Planck$+$WP data \cite{Planckdata,Planck2015}. In the former case, one has $n_s=0.9603\pm0.0073$. A quick look at our numerics enables us to conclude that they lie in a an acceptable interval, within the $1\sigma$ confidence level.

On the other side, we get:
\begin{equation}
|r|\sim 0.0044\,,\label{uno8}
\end{equation}
which is compatible with the Planck constraint: $r<0.11$ \cite{Planckdata,Planck2015} at $2\sigma$ confidence level.

With those results in mind, the amplitude of the primordial power spectrum does not depart significatively from the expected values. Hence,  one concludes that our approach is well defined, even at the level of early time cosmology confirming to be suitable to  describe the universe dynamics at inflationary  time.


\section{Conclusions}

In this paper, we considered the imprint of quantum cosmology as a concrete prerogative to explain the universe dynamics. in this perspective, we adopted the  Von Neumann entropy into the context of general relativity, assuming that it can be considered as a source for the dark energy evolution. In particular, the basic assumption is that   dark effects propagate  up to our time starting from primordial quantum counterparts. In particular, we considered  the \textit{entangled state ansatz} assuming  that any  two epochs of the cosmic evolution are correlated  through the entanglement. In so doing, writing up the density matrix for  simple pairs of entangled wave functions, associated to distinct epochs, we are able to characterize the Von Neumann entropy as the  quantum information source driving the cosmic speed up. So that, in our treatment we assume that early phases of the universe evolution were described in terms of entangled quantum states, whose net effect is producing a \emph{thermodynamic source} described by  entropy contributions. We do not assume any microscopic landscape for characterizing $S$. In so doing, we frame the entanglement measure in terms of $S$, showing a coarse grained mechanism behind the  acceleration of the Hubble fluid. In such a picture, we even considered the linear entropy as a limiting case of the Von Neumann entropy for the specific case in which it is linearly expanded around the chemical potential $\mu[\rho]\sim\rho$. We required that late-time effects are comparable with those at early phases, having that the entropies, contributing to the net densities, evolve in a way compatible with observations. The adiabatic evolution can be investigated in such an approach   by defining a  {\it dark temperature} that matches information entropy   with standard thermodynamics. We interpreted it as the temperature due to quantum information, associated to the dark constituents of the universe. Its magnitude turns out to be negligibly small, leading to the gravitationally repulsive effects of dark energy. We even showed that the entropy density may be easily compared with an information entropy, by comparing $S$ with the given probability of forming structures.  We thus demonstrated that the cosmological constant $\Lambda$ can be  achieved in the simplest case of adiabatic evolution. This may be viewed as the possibility to \emph{fuel} the $\Lambda$ term by means of the information density, defined as a measure of the entanglement-degrees related to the thermodynamical state of the universe. Analogously, we showed that inflation is also recovered in this approach. Specifically, we defined an effective potential in terms of the scale factor $a(t)$, interpreting the inflaton as a field related to the entanglement processes. The numerical values of $n_s$, $r$ and $\Delta^2_{\cal R}$ are compatible with current observations as well as the constraints used for late-time cosmology.
In future works, we will develop the approach taking into account observational data at early and late epochs.

\section*{Acknowledgements}

S.C. acknowledges the support of  INFN ({\it iniziativa specifica} QGSKY). This paper is  based upon work from COST action
 CA15117 (CANTATA), supported by COST (European Cooperation in Science and Technology).

\end{document}